\documentclass[prl,twocolumn,showpacs]{revtex4}
\usepackage{graphicx}
\usepackage{amsmath}
\begin{document}
%
\title{Self-Focussing Dynamics in Monopolarly Charged Suspensions}
\author{Stephan~M.~Dammer and  Dietrich~E.~Wolf}
\affiliation{Institut f\"ur Physik,
  Universit{\"a}t Duisburg-Essen, 47048 Duisburg, Germany}
\date{\today}
\begin{abstract}
  The Smoluchowski equation for irreversible aggregation in
  suspensions of equally charged particles is studied. 
  Accumulation of charges during the
  aggregation process leads to a crossover from power law to
  sub-logarithmic cluster growth at a characteristic time and cluster
  size. For larger times the suspension is usually called
  stable, although aggregation still proceeds slowly. 
  In this regime
  the size distribution evolves towards a {\em universal} scaling form,
  independent of {\em any} parameter included in the theory. The
  relative width falls off to a universal value $\sigma
  _{\rm r}^{\infty}{\approx}0.2017$ that is much smaller than in the
  uncharged case. We conjecture that $\sigma _{\rm r}^{\infty}$ is a lower
  bound for the asymptotic relative width for all physical
  systems with irreversible aggregation.    
\end{abstract}
\pacs{05.40.-a,05.65.+b,45.50.-j}
\maketitle
In many fields of science irreversible aggregation (or coagulation,
agglomeration, clustering) phenomena are important.
Examples include aerosol coalescence, polymerization or gelation (see, e.g.,
\cite{Ernst,Lee,Ivlev,Friedlander,Fuchs} and references therein). Such
systems have been described using Smoluchowski's
coagulation equation and extensive analytical and numerical studies
have been carried out for uncharged systems~\cite{Ernst,Lee}. 
In spite of the fact that charges are present in a lot of aggregating
systems, much less is known for the charged case. In this Letter
we investigate the influence of charges on the aggregation process.
In contrast to previous work which studied aggregation between
oppositely charged particles~\cite{Ivlev} we investigate the aggregation dynamics
of equally charged particles with the focus on suspensions.
A practical example is given in
\cite{Jochen}, where coating of medical drug particles with
nano-particles is investigated, in order to make them
inhalable. As an intermediate step it is desirable to keep 
each powder component in suspension separately, controlling its aggregation
by charging the particles equally. They collide due to Brownian motion, and strong van der
Waals forces result in irreversible aggregation (Brownian
coagulation). However, the results we derive are valid for more
general systems, as well. Asymptotically neither the solvent nor the diffusion 
properties are important, as long as particles may have high kinetic
energy, albeit as rare events, and 
charges accumulate in the aggregation process.

Usually a suspension is regarded as stable with respect to the aggregation of
equally charged particles, if the particle diameter $d$ is small
compared to the Bjerrum length $\ell_{\rm B}{=}q^2/4\pi\epsilon_0\epsilon _{\rm r}
k_{\rm B}T$, where $q$ is the charge of the particles, $T$ and $\epsilon _{\rm r}$ are the temperature
and the relative dielectric constant of the carrying fluid, $k_{\rm B}$ is
Boltzmann's constant and $\epsilon_0$ is the dielectric constant of
vacuum. The reason is that for $d{<}\ell_{\rm B}$
the energy barrier for bringing two particles
into contact is larger than the thermal energy. 

Consider the case, where a suspension is initially unstable (as
in~\cite{Jochen}), which means that initially aggregation
events are frequent. We are going to show that charge accumulation 
eventually leads to a crossover to slow aggregation 
at characteristic values of time $t_c$ and cluster size $s_c$. 
Above $t_c$
the Bjerrum length becomes larger than the aggregate diameter, so
that one would regard the suspension as stable for later times. 
However, the cluster size distribution keeps evolving slowly and
approaches a {\em universal} self-similar form which is independent of
{\em any} parameter included in the theory. A
particular important consequence of this is that the relative width of the
cluster size distribution $\sigma _{\rm r}$ starts decreasing at $t_c$ and
asymptotically reaches a
universal value $\sigma _{\rm r}^{\infty}{\approx}0.2017$. In this sense, the cluster size
distribution becomes narrower and we refer to this effect as charge-induced
{\em self-focussing}. For Brownian coagulation the value $\sigma _{\rm
  r}^{\infty}{\approx}0.2017$ must be compared to $\sigma _{\rm r}{\approx}1$ for the
uncharged case. We conjecture that $\sigma _{\rm r}^{\infty}$ is a lower
bound for the asymptotic value of $\sigma _{\rm r}$ for all physical
systems with irreversible aggregation. 
Our results are valid as long as aggregation
events occur frequently compared to dissociation of clusters
(neglected in the following).

We investigate Smoluchowski's coagulation equation for
monodisperse initial conditions, i.e. 
initially all particles have the
same mass $m^*$ and radius $a^*$ and carry the
same charge $q^*$. As mass and charge of aggregates are
proportional to each other in this case, one index $i$ is sufficient in
the coagulation equation 
\begin{equation}\label{coagulation_eq}
  \frac{dn_i(t)}{dt}=\frac{1}{2}\sum
  _{j+k=i}R_{jk}n_j(t)n_k(t)-n_i(t)\sum _{j=1}^{\infty}R_{ij}n_j(t).
\end{equation}
$n_i(t)$ denotes the number density at time $t$ of clusters of mass
$m_i{=}im^*$,
charge $q_i{=}iq^*$ and radius $a_i{=}i^\alpha a^*$. 
$1/\alpha$ is the fractal dimension of the aggregates. If
they are spherical (e.g., coagulating droplets) one has
$\alpha{=}1/3$. More generally one may assume that $1{>}\alpha{>}1/3$
for fluffier aggregates. The initial conditions are $n_1(t{=}0){=}1$
and $n_i(t{=}0){=}0\text{ for }i{>}1$.  $R_{ij}$ is the rate coefficient 
(coagulation kernel) for mergers between clusters with indices $i$ and
$j$. Eq.~(\ref{coagulation_eq}) neglects spatial fluctuations
and correlations. In this sense it is a mean field equation. We checked, that
a Brownian dynamics simulation of aggregating charged spheres in three
dimensions is in accordance with the mean field description~\cite{book}.

The rate coefficients $R_{ij}$ depend on the particular physical
system under consideration. In suspensions clusters of mass $m_i$
diffuse with diffusion constant $D_i$. Usually the van der Waals
interaction is represented as a sticky contact force, leading to
irreversible aggregation. Therefore, aggregation with a cluster of mass
$m_j$ occurs when the distance between the clusters equals
$a_i{+}a_j$ (Brownian coagulation). A finite range of the van der Waals interaction would
basically increase this 'contact distance' by a constant, which becomes less
and less important for growing clusters. For Brownian coagulation $R_{ij}$ can
be calculated by solving the diffusion equation in the presence
of an absorbing sphere~\cite{Friedlander,Fuchs} leading to \mbox{$R_{ij}{=}4\pi
  (a_{i}+a_j)(D_{i}+D_j){\equiv}r_{ij}$}. If Einstein's formula applies, one can insert
$D_i{=}k_{\rm B}T/6\pi\eta a_i$, with the viscosity $\eta$ of the carrying
fluid. This leads to 
\begin{equation}\label{brown_rate_1}
r_{ij}=\frac{2k_{\rm B}T}{3\eta}(i^{\alpha}+j^{\alpha})(i^{-\alpha}+j^{-\alpha})\ .
\end{equation}

For most {\em uncharged} coagulating systems discussed in the literature the
coagulation kernels are homogeneous functions of degree $\lambda$,
i.e. $R_{bi\,bj}{=}b^\lambda R_{ij}$, with
$\lambda{\leq}2$~\cite{Ernst,Friedlander}.
Eq.~(\ref{brown_rate_1}), for instance, corresponds to
$\lambda{=}0$. It turns out that $\lambda$ is the most important parameter for
the uncharged case. Depending on its value, the solutions of the
coagulation equation are qualitatively different. For $\lambda{\leq}1$
the cluster size distribution decays exponentially for large sizes at all times. In the
scaling limit the solutions are self-similar and (for $\lambda{<}1$)
have the form 
\begin{equation}\label{sim_sol}
n_i(t)=s(t)^{-2}\phi\left(i/s(t)\right)\text{ with }\,s(t)\sim
t^\frac{1}{1-\lambda}\ .
\end{equation} 
The scaling function $\phi$ depends on the details of the
underlying coagulation kernels. The average size $s(t)$ is a measure for the
average number of primary particles per cluster. For
$2{\geq}\lambda{>}1$ the cluster size distribution
develops a power-law tail in finite time $\tilde{t}$ at 
large sizes that violates mass conservation~\cite{Ernst,ziff} ({\em
  runaway growth}), which is associated with the occurrence of a {\em
  gelation transition}. 

A derivation similar to the uncharged case leads to the coagulation
kernel for Brownian coagulation of {\em charged} clusters~\cite{Fuchs}  
\begin{equation}\label{charged}
\begin{array}{lcl}
R_{ij}&=&W_{ij}r_{ij}\ ,\
W_{ij}=\kappa_{ij}/\left[\exp(\kappa_{ij})-1\right]\ ,\vspace{2mm}\\
\kappa_{ij}&=&q_iq_j/\left[4\pi\epsilon _0\epsilon _{\rm r}(a_i+a_j)k_{\rm B}T\right]
\end{array}
\end{equation}
where $r_{ij}$ is the coagulation kernel for uncharged
clusters, Eq.~(\ref{brown_rate_1}). It was assumed that at contact of the clusters the
charges are separated by a distance $a_i{+}a_j$.  

The important feature of $W_{ij}$ is that collisions of strongly
charged particles ($\kappa _{ij}{\gg}1$) are exponentially
suppressed. We are going to show that $W_{ij}{=}\exp(-\kappa_{ij})$
(which is found in charged granular gases~\cite{Tim}) leads to the same
asymptotic results.

The purpose of this Letter is to study the behavior
of Eq.~(\ref{coagulation_eq}), depending on the parameters $\lambda$,
$\alpha$ and $k{=}\sqrt{{q^*}^2/(4\pi\epsilon _0\epsilon _{\rm r}
  a^*k_{\rm B}T)}$, for rate coefficients that obey
\begin{eqnarray}
R_{ij}&=&(i^{\alpha}+j^{\alpha})(i^{-\alpha}+j^{-\alpha})\,\frac{\kappa_{ij}}{\exp(\kappa_{ij})-1}
\label{rate_prototype_1}\\
{\rm or} \quad R_{ij}&=&(ij)^{\lambda /2}\,\exp(-\kappa_{ij})\label{rate_prototype_2}\\
{\rm with}\quad\kappa_{ij}&=&\frac{k^2ij}{(i^\alpha+j^\alpha)}\label{kappa}\ .
\end{eqnarray}
These expressions are motivated by the observations presented
above. $\kappa_{ij}$ is proportional to the Coulomb energy
at contact of the clusters divided by a temperature-like
variable. Eq.~(\ref{rate_prototype_1}) is  the dimensionless
coagulation kernel 
for Brownian coagulation of charged clusters, Eq.~(\ref{charged}),
with $2k_{\rm B}T/3\eta{=}1$. Whereas this corresponds to $\lambda{=}0$ as
was pointed out above, we investigate the influence of
$\lambda{\neq}0$ using the general rate coefficient
Eq.~(\ref{rate_prototype_2}). 

Let the system initially be unstable ($k^2{\ll}1$) which means that in the
beginning the exponential terms in Eqs.~(\ref{rate_prototype_1},\ref{rate_prototype_2}) are
close to unity. The system basically behaves as if it was
uncharged. This changes as soon as values of $\kappa_{ij}{\approx} 1$
become important so that the exponential terms must be taken into
account. Then further aggregation is exponentially suppressed and cluster
growth becomes very slow. This happens when the average cluster size $s$ approaches a
characteristic value $s_c$ given by $\kappa_{s_c s_c}{\approx} 1$. 
Note that even systems which in the uncharged case exhibit runaway growth,
i.e.~$1{<}\lambda{\leq}2$, eventually cross over to slow
aggregation. 
According to Eq.~(\ref{sim_sol}) the crossover time $t_c$ is given by
$s_c{\sim}t_c^\frac{1}{1-\lambda}$ for $\lambda{<}1$, leading to 
\begin{eqnarray}
&s_c\approx k^{-\frac{2}{2-\alpha}}\ ,\ t_c\approx k^{-\frac{2-2\lambda}{2-\alpha}}\, t^*\ \ \text{for}\ \lambda <
1\label{characteristic}\ \ ,
\end{eqnarray}        
where $t^*$ is the appropriate time unit. Note that the expression for
$s_c$ is also valid for $\lambda{\geq}1$. Though similar arguments may be
applied for $t_c$ and $\lambda{\geq}1$ we do
not discuss this case here.
\begin{figure}[t]
\includegraphics[width=85mm]{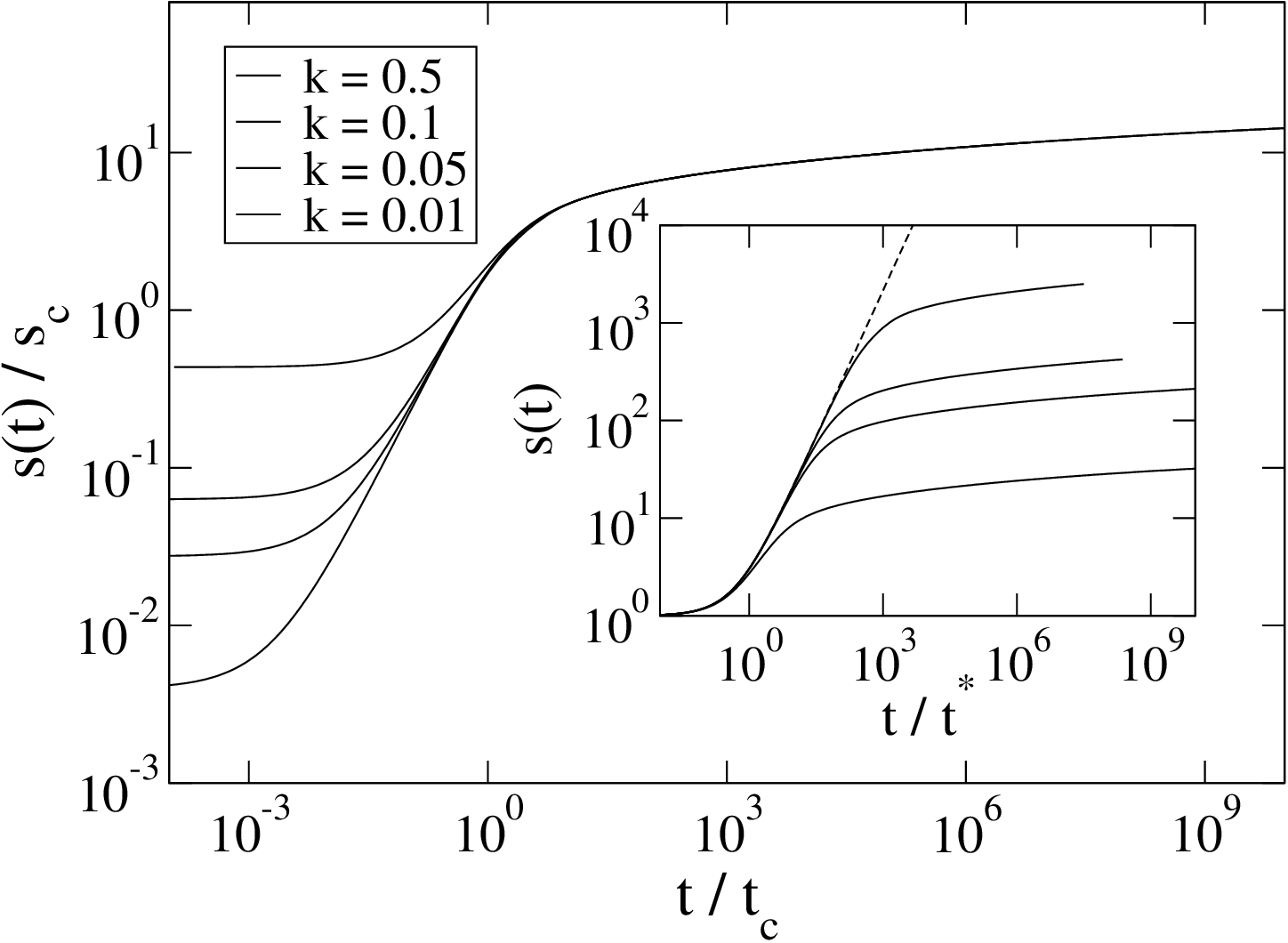}
\caption{
\label{M}
The average size $s(t)$ scaled with $s_c$ vs. time $t$ scaled with $t_c$ for
different values of $k$. The inset shows the original data, and the dashed line
is the behavior for an uncharged system ($k{=}0$). 
During integration the time step could be increased, since the rates for
mergers become small in the stable regime. Thus,
$t/t^*$ is larger than the number of time steps.  
}
\end{figure}

The rate equations (\ref{coagulation_eq}) were solved numerically for
various choices of the coagulation kernels $R_{ij}$ and different
values of $k$. Fig.~\ref{M} shows data for Brownian coagulation of
charged spheres (Eq.~(\ref{rate_prototype_1}) with $\alpha{=}1/3$). 
The average size $s(t)$ was calculated as
$s{=}M_1/M_0$ with the moments defined as $M_l{=}\sum _ii^ln_i$. 
$M_1$ is proportional to the constant total mass of the system and was
chosen equal 1. $s_c$ and $t_c$ were determined according to
Eq.~(\ref{characteristic}). 
Fig.~\ref{M} clearly shows the crossover from fast power-law
cluster growth to slow aggregation. Moreover,
Eq.~(\ref{characteristic}) is confirmed by the data
collapse. In addition, we checked rate coefficients obeying
Eq.~(\ref{rate_prototype_2}) with different 
values of $\lambda$ and $\alpha$ and found in all cases agreement with
Eq.~(\ref{characteristic}), as well. 

Comparison of the dimensionless rate (\ref{rate_prototype_1}) with
Eq.~(\ref{brown_rate_1}) shows that 
the time unit is $t^*{=}\frac{3\eta}{\tilde{n}2k_{\rm B}T}$, where
$\tilde{n}$ is the initial number density of primary particles.  
As an example, we compute the physical time scale $t^*$ for possible realistic
parameters. With $\eta{=}10^{-4}{\rm Pa\,s}$, $T{=}300{\rm K}$,
$a^*{=}1{\rm \mu m}$ and volume fraction $\nu{=}10^{-3}$ ($\tilde{n}{\approx}2{\cdot}10^5{\rm mm}^{-3}$) one obtains 
$t^*{\approx}150{\rm s}$. Assuming $\epsilon _{\rm r}{\approx}1$,
charging with a single elementary charge on each primary particle
corresponds to a value $k{\approx}0.2$. For these parameters one finds
with the dimensionless results of Fig.~\ref{M} that the crossover to slow aggregation
occurs within hours.  

\begin{figure}
\includegraphics[width=85mm]{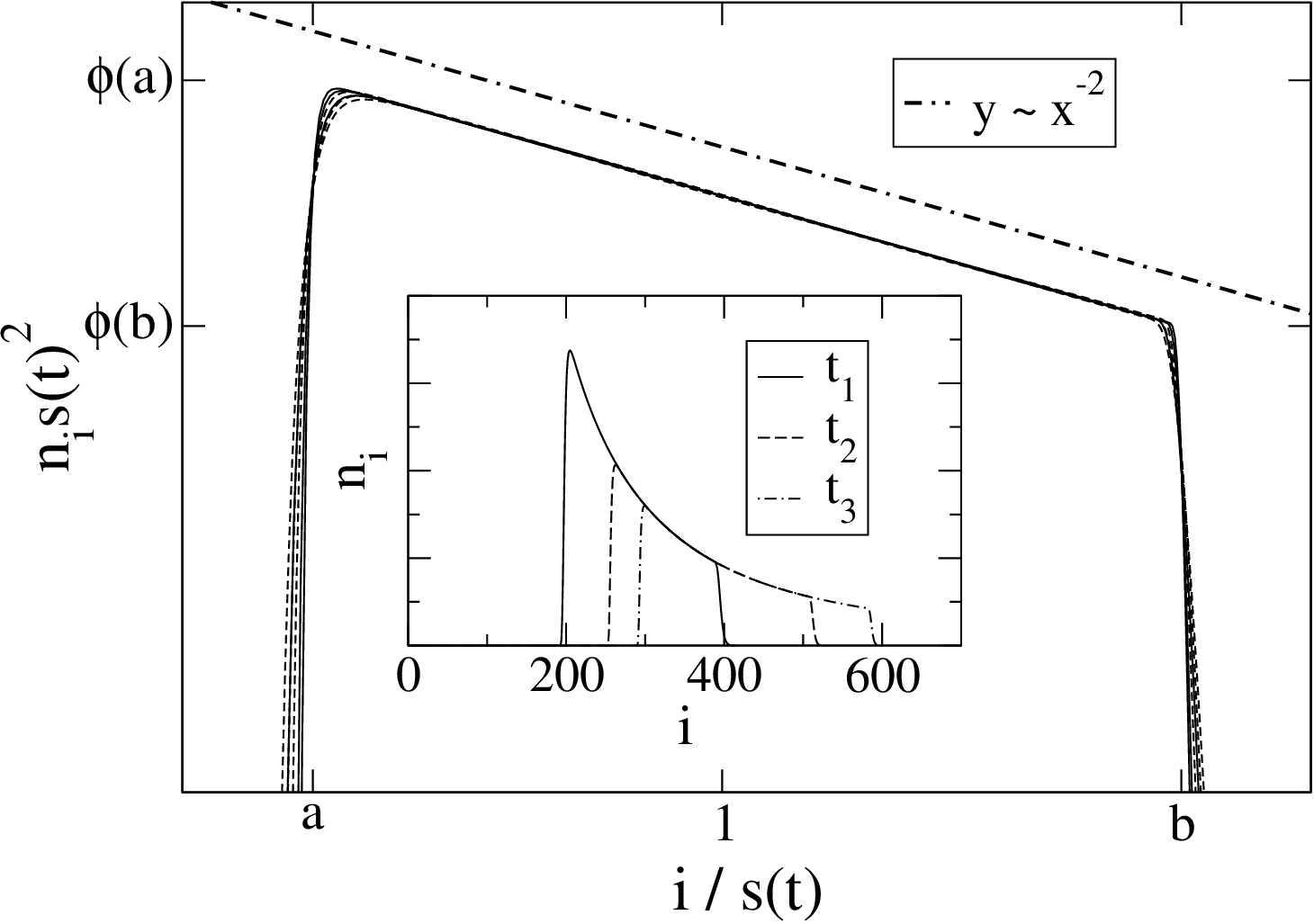}
\caption{
\label{n}
Double logarithmic plot of the scaled cluster size distribution for different
times and choices of the coagulation kernels. The inset shows original
data. 
}
\end{figure}

In the unstable regime the system behaves basically as in the
uncharged case and the cluster size distribution is approximately given by
Eq.~(\ref{sim_sol}). Fig.~\ref{n} displays a double logarithmic plot
of the scaled size distribution in the {\em stable} regime for three different times and two different
choices of the rate coefficients. The coagulation kernel for the dashed lines obeys
Eq.~(\ref{rate_prototype_1}) with
$k{=}0.1$ and $\alpha{=}1/3$ (Brownian coagulation of charged spheres).
Solid lines correspond to Eq.~(\ref{rate_prototype_2}) with
$k{=}0.05$, $\alpha{=}0$ and $\lambda{=}{-}2$. The
inset shows the latter data unscaled ($t_1{<}t_2{<}t_3$). 
The parameter $\alpha{=}0$ corresponds to the limiting case, where the
Coulomb repulsion of large clusters becomes independent of the cluster radius.
For comparison the dot-dashed line shows a decay $\sim x^{-2}$. The labels $a$, $b$, $\phi (a)$ and
$\phi(b)$ mark analytical expressions for the asymptotic edges of the
distribution and the corresponding heights, which will be derived in the
following.
It can be seen that for monopolarly charged systems the cluster
size distribution asymptotically  converges to a self-similar solution analogous to 
Eq.~(\ref{sim_sol}). Furthermore, Fig.~\ref{n} strongly supports the claim that
$\phi (x)$ is independent of the details of the underlying coagulation
kernel. The data suggest that the scaling function $\phi (x)$ decays
as $\phi (x){\sim}x^{-2}$ between sharp edges and is zero outside. As can be
observed in the inset of Fig.~\ref{n}, a consequence of this is that $n_i$ is
independent of the average size $s$ between the edges of the size
distribution. Based on the numerical data we assume
\begin{eqnarray}
  \phi(x)&=&\begin{cases} 
\phi _0\,x^{-2} & \text{ for } x\in [a,b] \\ 
0 & \text{ for } x\not\in [a,b]\\
\end{cases}
\label{phi}
\end{eqnarray}
Applying Eq.~(\ref{phi}) one finds $M_0{=}(\phi _0/s)\,(1/a{-}1/b)$
and $M_1{=}1{=}\phi_0\ln(b/a)$, together with $s{=}M_1{/}M_0$ leading
to 
\begin{equation}\label{ab_1}
\ln (b/a) = [1/a-1/b]\ .
\end{equation} 

The characteristic feature of the coagulation rates for monopolarly charged
systems is that (in the stable regime) coagulation is exponentially
suppressed with increasing cluster sizes. This feature does not depend
on $\lambda$, $k$, and $\alpha$ or whether one applies
Eq.~(\ref{rate_prototype_1}) or Eq.~(\ref{rate_prototype_2}). The
exponential suppression causes the rates for coagulation of
clusters located at the left edge of the size distribution to be much
larger than rates for any other coagulation events. Thus, clusters at the
left edge with index $i{\in}[as,as{+}\Delta i]$ coagulate, and mass
  conservation requires that this leads to an
increase of the number of clusters with $i{\in}[2as,2as{+}2\Delta
  i]$. Hence, we expect that asymptotically $b{=}2a$. Between the left
  and the right edge of the size distribution $n_i$ is essentially not
  changed. The scaling function Eq.~(\ref{phi}) is the only
  possibility to guarantee such a behavior. Inserting $b{=}2a$ into
  Eq.~(\ref{ab_1}) leads to 
\begin{equation}\label{result}
a{=}\frac{1}{2\ln{2}}\ ,\ b{=}\frac{1}{\ln{2}}\ ,\ \phi
_0{=}\frac{1}{\ln{2}}
\end{equation}
in good agreement with the numerical data, Fig.~\ref{n}.

\begin{figure}
\includegraphics[width=85mm]{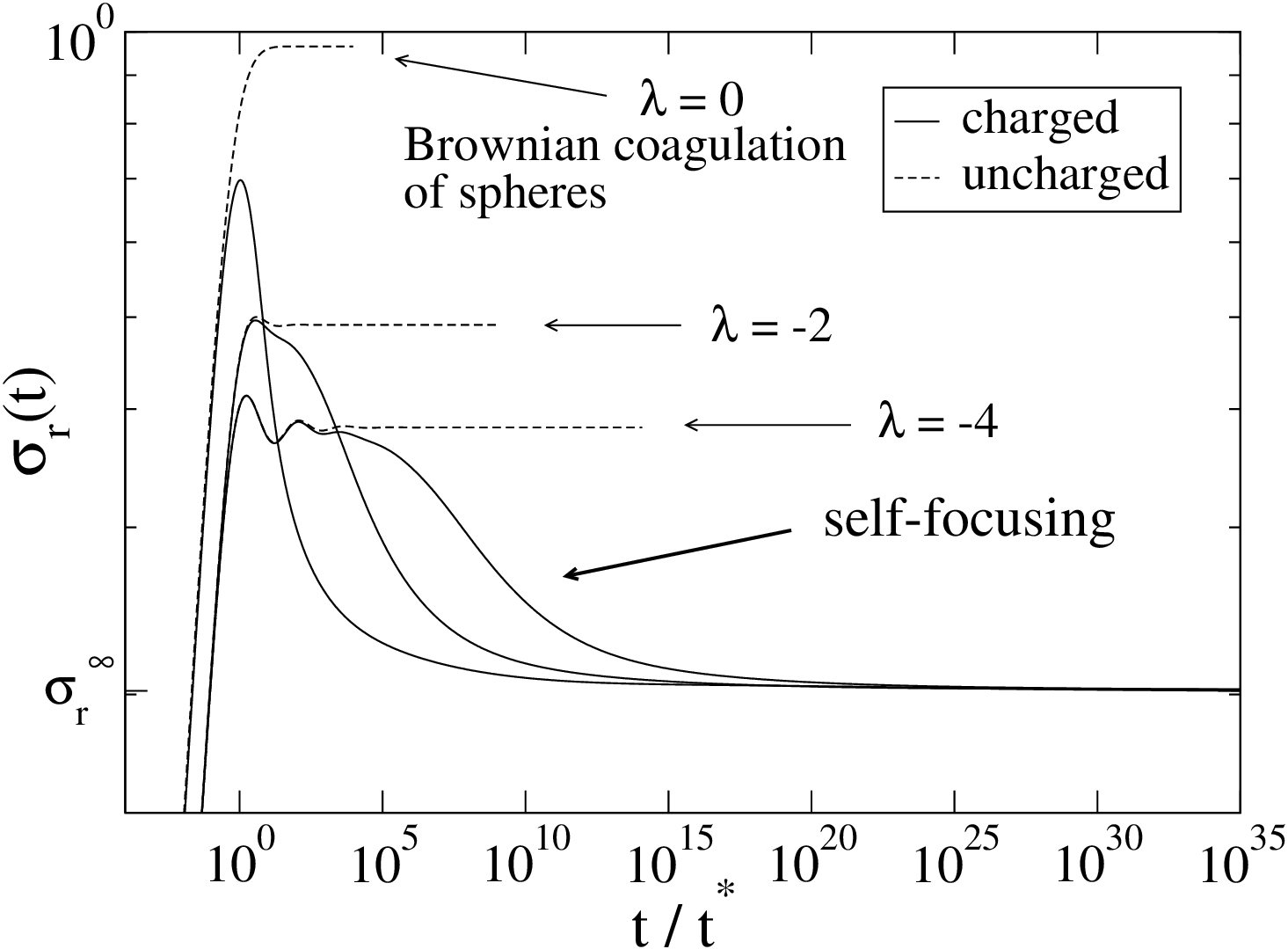}
\caption{
\label{s}
Relative width $\sigma _{\rm r}$ of the cluster size distribution for
monopolarly charged and corresponding uncharged
systems. $k=(0.5,0.1,0.05)$ for $\lambda = (0,-2,-4)$.}
\end{figure}

An important consequence of the universal form
Eqs.~(\ref{phi},\ref{result}) is that the relative width of the cluster size
distribution $\sigma _{\rm r}{=}\sqrt{(M_2M_0/M_1^2)-1}$ approaches a
universal value $\sigma _{\rm r}^{\infty}$ with 
\begin{equation}\label{sigma_inf}
\sigma_{\rm r}\ \longrightarrow\ \sigma_{\rm r}^{\infty}=\sqrt{1/[2(\ln{2})^2]-1}\approx 0.2017\ .
\end{equation} 
Fig.~\ref{s} shows numerical data for $\sigma _{\rm r}$ with $\lambda {=}
(0,{-}2,{-}4)$ for monopolarly charged particles (solid lines) in
comparison to the corresponding uncharged cases
(dashed lines). 
$\sigma_{\rm r}^{\infty}$ is considerably
smaller than the relative width for any physical system we found in
the literature. 

For an initially narrow cluster size distribution 
the relative width first grows similar to the uncharged situation until 
time $t_c$, when it starts decreasing again, induced by
the exponential suppression of further aggregation. We refer to this as {\em
  self-focussing}. The example discussed above, with
$t^*{\approx}150{\rm s}$, shows that this self-focussing effect can occur within
an experimentally accessible time.  

\begin{figure}[h]
\includegraphics[width=75mm]{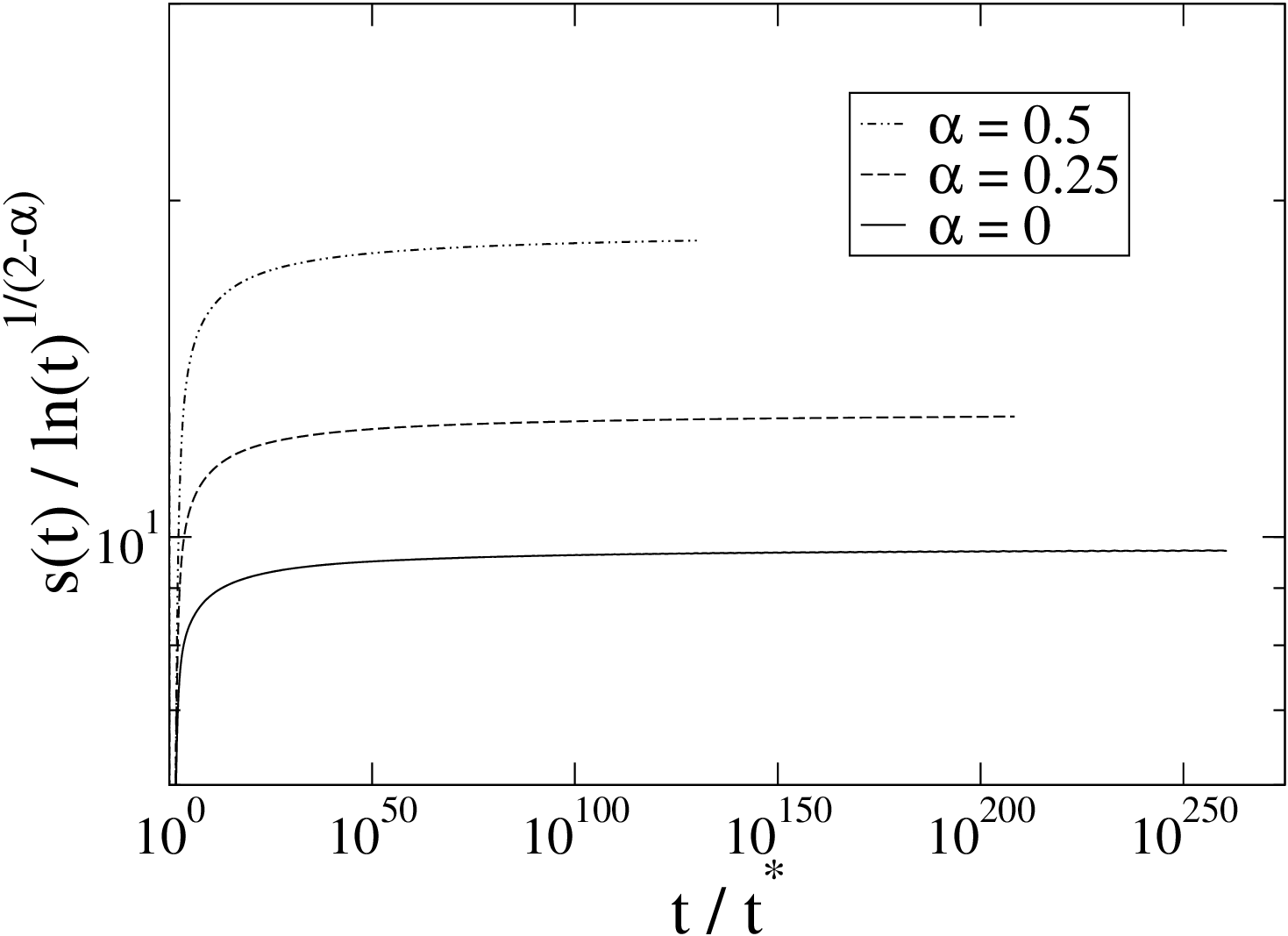}
\caption{
\label{s_a}
$s(t)$ times $\left[\ln
  (t)\right]^{-\frac{1}{2-\alpha}}$ for different values of $\alpha$. 
}
\end{figure}

In the following, we specify what kind of time dependence the average
size $s$ exhibits in the stable regime. For this purpose we calculate
the time derivative $\frac{d}{dt}M_0=\sum _i\frac{d}{dt}n_i(t)$. According to the previous
results, in the stable regime $\frac{d}{dt}M_0$ reduces to a
transfer from the left edge of the size distribution ($i{\approx}as$)
to the right edge ($i{\approx}2as$). This behavior
together with Eq.~(\ref{phi}) leads to $\frac{d}{dt}M_0 =
-\frac{1}{2}\,R_{as\,as}\,\phi
^2_0/(as)^4$. In the stable regime the coagulation
kernels asymptotically obey $R_{y,y}\sim\exp (-k^2y^{2-\alpha})$. Using
$s=M_1/M_0$ we arrive at
\begin{equation}\label{diff_s}
\frac{ds(t)}{dt}\sim\frac{1}{s(t)^2}\,e^{-Ks(t)^{2-\alpha}}
\text{ with } K=k^2a^{2-\alpha}\ . 
\end{equation}
This results in the time interval
$t_2-t_1\approx I(s_2)-I(s_1)$ in which the average cluster size
increases from $s_1$ to $s_2$, with $I(y)=\int
_1^{y} dx\,x^2\exp (Kx^{2-\alpha})$. Application of
$I(y)\approx y^{\gamma}\exp (Ky^{2-\alpha})$ for large $y{\gg}1$, with
the exponent $\gamma$ depending on
$\alpha$, leads to
$s(t)\sim (\ln{t}-\gamma\ln{s(t)})^{\frac{1}{2-\alpha}}$. Thus, we can
conclude that asymptotically 
\begin{equation}\label{asy_s}
s(t)\sim (\ln{t})^{\frac{1}{2-\alpha}}\ .
\end{equation} 
A similar long-time behavior was observed
  in~\cite{Tim,poschel,ko}. Fig.~\ref{s_a} shows numerical data for $R_{ij}$ obeying
  Eq.~(\ref{rate_prototype_2}) with $\lambda {=}0$, $k{=}0.2$ and
  different values of $\alpha$, in accordance with the predicted
  asymptotic behavior in Eq.~(\ref{asy_s}).  

In conclusion, we investigated Smoluchowski's coagulation equation for
monopolarly charged particles. Charge accumulation leads to a
crossover from power-law to sub-logarithmic cluster
growth. Asymptotically the cluster size distribution approaches a {\em universal}
scaling form, independent of {\em any} parameter included. This
implies a {\em self-focussing} of the relative width of the
distribution. A first experimental check is encouraging~\cite{wirth}. 

We thank H.~Kundsen, J.~Werth and H.~Hinrichsen for fruitful
discussions and the DFG for support within the program ``Verhalten granularer
Medien'', project Hi/744.

\end{document}